# Structure-selective operando x-ray spectroscopy

Daniel Weinstock[1], Hayley S. Hirsh[2], Oleg Yu. Gorobtsov[1], Minghao Zhang[2], Jason Huang[1], Ryan Bouck[1], Jacob P. C. Ruff[3], Y. Shirley Meng[2], and Andrej Singer[1]

[1]*Department of Materials Science and Engineering, Cornell University, Ithaca, NY.*

[2]*Department of NanoEngineering, University of California San Diego, La Jolla, CA.*

[3]*Cornell High Energy Synchrotron Source, Cornell University, Ithaca, NY.*

**The relationship between charge and structure dictates the properties of electrochemical systems. For example, reversible Na-ion intercalation – a low-cost alternative to Li-ion technology – often induces detrimental structural phase transformations coupled with charge compensation reactions[1,2]. However, little is known about the underpinning charge-structure mechanisms because the reduction-oxidation (redox) reactions within coexisting structural phases have so far eluded direct operando investigation. Here, we distinguish x-ray spectra of individual crystalline phases operando during a redox-induced phase transformation in P2-$Na_{2/3}Ni_{1/3}Mn_{2/3}O_2$ – an archetypal layered oxide for sodium-ion batteries[1,3]. We measure the resonant elastic scattering on the Bragg reflection corresponding to the P2-phase lattice spacing. These resonant spectra become static midway through the sodium extraction in an operando coin cell, while the overall sodium extraction proceeds as evidenced by the X-ray absorption averaging over all electrochemically active Ni atoms[4,5]. The stop of redox activity in the P2-structure signifies its inability to host $Ni^{4+}$ ions. The coincident emergence of the O2-structure reveals the rigid link between the local redox and the long-range order during the phase transformation. The structure-selective x-ray spectroscopy thus opens a powerful avenue for resolving the dynamic chemistry of different structural phases in multi-phase electrochemical systems.**

Improving the understanding of reduction-oxidation (redox) chemistry in solid-state systems is essential for technological applications such as energy storage[6] and conversion.[7] Battery technologies are of increasing interest for grid storage because they are highly modular, have fast start-up/wind-down times, and store wind and solar energy locally. Sodium-ion batteries (NIBs) are promising alternatives to lithium-ion batteries (LIBs) due to their lower cost of energy and the natural elemental abundance of sodium.[2,8,9] Similar to their LIB analogs,[10] the most promising NIB electrode materials are layered oxide materials ($Na_xMO_2$ where $x \leq 1$ and M is a transition metal), in which sodium ions ($Na^+$) intercalate between neighboring layers of $MO_6$.[11,12] The transition-metal cations compensate the electric charge for the dynamic sodium-ion concentration through redox reactions, and together with the oxygen framework provide structural stability.[11] The electrochemical behavior of NIB layered oxide cathodes is inherently linked to the species and oxidation state of the transition metal in the host structure, leading to variances in the coordination of sodium ions, (de)sodiation kinetics, operation voltages, and the batteries' lifetimes.[2] Despite significant research efforts over the last decades, layered NIBs materials still lack desired energy densities and durability, both linked to the prevalence of structural phase transformations during operation and our lack of detailed understanding of the chemistry-charge -structure relationships in these systems.[2,12]

Many techniques exist to determine the oxidation state of the transition metal cations in electrochemical systems. Yet, few can provide spatial resolution to distinguish between structural phases coexisting at the nanoscale, for example, within sub-micron particles used in technologically relevant electrode assemblies. X-ray absorption spectroscopy (XAS) is well-established for studying electrochemical systems.[13] It allows for the determination of the oxidation state of transition metals in catalysts and intercalation hosts, and in combination with voltammetry, can determine which species are electrochemically active.[4,5,14] XAS combined with x-ray microscopy can provide local information on chemical processes in secondary electrode particles[15] and large primary particles;[16] nevertheless, the resolution is still limited for studying chemistry in primary particles in technologically relevant sizes (<1 $\mu$m) operando, and absorption microscopy experiments lack the direct access to the crystalline long-range order. Electron energy loss spectroscopy (EELS) combined with scanning transmission electron microscopy (STEM) can probe oxidation states with atomic resolution.[17,18] However, the small penetration depth of electrons requires ex-situ measurements or in-situ analogs that may not fully recreate the real operating conditions of electrochemical systems.[19,20]

Resonant elastic x-ray scattering (REXS) combines diffraction with spectroscopy. In REXS, one records the intensity of a Bragg reflection while scanning the x-ray energy across an absorption edge of the atomic species under study. Conventional REXS has been applied to large crystals in condensed matter physics [21–24] and systems with larger-range ordering such as polymers and liquid crystals using soft x-rays (RSoXS). [25,26] Diffraction anomalous fine structure (DAFS) emerged as a tool to conduct site-selective spectroscopy; however, its application to large single crystals and ex-situ systems coupled with the demanding signal requirements limited its impact on materials science[24,27–29]. Here we develop operando resonant elastic x-ray scattering (oREXS), which enables us to isolate x-ray spectra of a specific crystallographic phase coexisting with another crystallographic phase at the nanoscale during a structural phase transformation. We investigate the structural and chemical evolution of a model positive electrode material for sodium-ion batteries, P2-$Na_xNi_{1/3}Mn_{2/3}O_2$ (0<x<2/3). The material features the notorious P2-O2 structural phase transition prototypical for layered-oxide intercalation hosts for sodium ions. During the transient P2-O2 phase coexistence, the two phases have distinct Bragg diffraction peaks. We focus on the resonant (002) Bragg reflection corresponding to the lattice spacing of the P2-phase, and therefore collect x-ray spectra from the P2-phase only. By measuring around the Nickel (Ni) absorption edge, we find that the Ni-cations in the P2-type phase display a maximum oxidation state of $Ni^{3+}$. Rather than oxidizing further to $Ni^{4+}$ within the P2 structure, the material transitions to an O2-type structure with $Ni^{4+}$. Our results directly show that the oxidation state of the electrochemically active transition metal rigidly couples to the long-range order during the structural phase transformation.

Thomson scattering adequately describes light-matter interactions at x-ray photon energies far away from absorption edges: the atomic form factor is the Fourier transform of the electron density, $\rho(r)$, within the atom $f^0(Q) = \int \rho(r)e^{iQr}dr$, in units of the classical electron radius $r_0$, where $Q = (4\pi/\lambda) \cdot \sin(\theta)$ is the reciprocal space vector, $\lambda$ is the x-ray wavelength, and $\theta$ is the diffraction angle. Close to an absorption edge, the energy of the x-ray photon becomes essential, and both photoelectric absorption and resonant scattering processes contribute to the total scattering. Above resonance, an atom can absorb a photon by promoting an inner-shell electron to a higher energy level or the continuum (see Fig.1a). X-ray absorption provides information about the immediate environment of the probed atom. It is independent of the long-range order because the photon and its quantum mechanical phase vanish during absorption (see Fig. 1b). Therefore, in a system where the probed species exist in multiple environments with varying long-range order, the resulting absorption spectrum will be an average from all illuminated atoms. Close to the resonance, the atom can also coherently absorb and reemit a photon via an intermediate bound state (see Fig. 1c). The resonant scattering process maintains the quantum mechanical phase of the scattered photon because the scattering atom is indistinguishable from all identical atoms in a crystal (see Fig. 1d). Distinct Bragg peaks emerge in the reciprocal space, and REXS tuned to a specific Bragg peak measures the spectrum of that crystalline phase alone.

The energy-dependent atomic form factor accounting for the resonant scattering includes the real $f'(\omega)$ and imaginary $f''(\omega)$ dispersion corrections: $f(Q,\omega) = f^0(Q) + f'(\omega) + if''(\omega)$. Since the spatially constrained inner-shell electrons dominate the behavior of $f'(\omega)$ and $f''(\omega)$, the dispersion corrections can be considered Q-independent. The optical theorem allows determining $f''(\omega)$ from the experimentally found absorption cross-section, and Kramers-Kronig relations intrinsically connect $f'(\omega)$ with $f''(\omega)$.[30] For crystals, the scattering takes the form of the structure factor $F_{hkl}(Q,\omega) = \sum_j f_j(Q,\omega) e^{iQ \cdot r_j}$, where $f_j(Q,\omega)$ is the atomic form factor and $r_j$ are position in the unit cell of atom $j$. Figures 1e shows the imaginary dispersion correction $f''$ of a free-standing Ni-atom, and Figure 1f shows the intensity of the (002) diffraction peak as a function of the photon-energy for the P2-$Na_{2/3}Ni_{1/3}Mn_{2/3}O_2$ crystal structure around the Ni resonant edge, where we calculated $F_{hkl}(Q)$ by using all atoms in the unit cell and the dispersion corrections for a free-standing Ni-atom. Because $f'(\omega)$ and $f''(\omega)$ are tied by the Kramers-Kronig relations, both the absorption and resonant scattering display a spectroscopic feature at the resonant frequency. Importantly, the resonant scattering shows an energy dependence at a lower photon energy than the absorption – the latter is energy-independent until a discontinuous jump at $\omega=\omega_s$, and both relate to the same $\omega_s$ (shown with a dotted line at 8333 eV) – allowing one to distinguish the resonant scattering from absorption in the experiment. A shift in the energy levels (see Fig. 1a,c) – for example as a result of redox activity – causes a shift of the resonance frequency observed in both spectra.[30,31]

We combined operando x-ray absorption, diffraction, and resonant x-ray scattering to investigate the relation between the P2-O2 structural phase transformations and redox activity of Ni in the layered P2-$Na_xNi_{1/3}Mn_{2/3}O_2$ (0<x< 2/3) (NNMO) – a cathode material with a reversible capacity of 140 mAh/g for its first cycle.[1,3,32] We used an operando sodium-ion coin cell with a sodium metal anode, similar to operando lithium-ion coin cells.[33,34] We charged the cell with NNMO cathodes at a rate of C/10 during the synchrotron-based x-ray measurements (see Fig. S1 for the experimental setup, two independent coin cells were measured). Figure 2a shows the voltage profile during charging to 4.5 V. The voltage profile shows multiple plateaus corresponding to two-phase regions.[35] Non-resonant x-ray diffraction ($\hbar\omega$=15 keV) confirms the presence of multiple structural phase transformations (see Figure 2b,c), readily visible in the abrupt shifts of the Bragg peak during charge. Each Bragg peak corresponds to a different lattice constant, and Figures 2b and 2c display Bragg reflections corresponding to the O2 and P2 phases, respectively (see Fig. 2d for the schematics of the two structures). The diffraction intensity of the two peaks shows the formation of the O2 phase at the expense of the P2 phase. NNMO retains a P2 structure for 1/3<x<2/3, with two intermediate phases corresponding to x=1/2 and x=1/3 due to ordering within the Na layers. At x<1/3, a two-phase region of P2 and O2 begins and continues as all Na is removed from the cathode, in agreement with literature.[1,3,5] The intensity of the P2 (002) peak increases before the O2 phase begins to form, which is in agreement with calculations shown in Figure S6a, where the intensity increases with decreasing Na content. Once O2 starts to form, the P2 (002) peak decreases in intensity along with the increasing intensity of the O2 peak (see Figure S7).

We performed operando resonant x-ray measurements on the same cells by measuring both the transmission and resonant Bragg signal around the nickel K-edge at 8.33 keV. Figure 3a shows the intensity of the (002) peak for the P2 structure measured resonantly, and the starts and ends of subsequent energy scans are marked with vertical bars. Figure 3b shows XAS (top) and REXS (bottom) spectra taken at the pristine condition and after charging to 4.2 and 4.5 V. As calculated in Figure 1e, the absorption spectra show a sharp increase at the resonant frequency $\omega_s$. Figure 3c shows the position of the absorption edge and how it evolves, calculated as the energy at which the measured absorption increases by half of its maximum jump in intensity (see methods). The absorption edge shifts to higher energies (with a slope of 7.4 eV/mol extracted Na) during the entirety of the charge: we extract sodium ions with a constant current resulting in a continuous extraction of electrons from Ni. The energy shift corresponds to charging the Ni from 2+ (discharged, low energy) to 4+ (fully charged, high energy).

The (002) REXS spectra show a dip in intensity as a function of energy, reaching a minimum at a resonant photon energy of $\hbar\omega_s$ and then increasing afterward, as expected by the calculations in Figure 1f. We determined the positions of the dip in the (002) diffraction intensity of the resonant spectra by finding the energy at which the change in intensity was half the maximum change (see methods). Like the XAS edge, the oREXS edge shifts to higher energies during charging for the first half of the charge (2/3>x>1/3): the oREXS edge follows an almost identical slope to the XAS edge (7.0 eV/mol extracted Na), indicating that the charge compensation for Na-extraction solely comes from the Ni ions in the P2 phase. Importantly, though the slope is the same, the REXS spectra have a calculated position of about 3 eV lower than the position of the XAS spectra, showing that the dip in (002) intensity is not merely an artifact of the absorption of the scattered radiation transmitted through the electrode (see supplementary information). At approximately x=0.3, the edge position from (002) REXS spectra plateaus as the XAS spectra continue to shift to higher energy. Because the position of the edge corresponds to the oxidation state of the P2 structure only, oREXS reveals that the Ni ions in the P2 phase are no longer electrochemically active after at x=1/3 even though the battery continues to charge.

While XAS is sensitive to all Ni atoms present in the NNMO cathode, oREXS only measures the Ni atoms coordinated in the P2 crystal structure, which we select by tuning the diffraction geometry to the P2-(002) Bragg peak. The interruption of Ni oxidation in the P2 phase indicates that another electrochemically active structural phase must form to accommodate the further oxidation apparent from XAS. We argue that this phase is the O2 phase whose formation we see non-resonantly in Figure 2b. Since the plateau starts at approximately x=1/3, the maximum oxidation state of the Ni-ions in the P2 phase is $Ni^{3+}$, and further oxidation to $Ni^{4+}$ induces the phase transformation to O2 (shown schematically in Figure 3d). As Ni-cation is the only electrochemically active transition metal in the cathode, the oxidation state of the Ni changes exactly with the change in concentration of $Na^+$. Therefore, we conclude that the minimum concentration of sodium ions necessary to stabilize the P2 phase is approximately x = 1/3, experimentally confirming the expectations from theoretical thermodynamic calculations.[3] The agreement between the operando measurements with the theoretical prediction indicates negligible over-charging of the P2 phase, suggesting a smaller energy barrier for the nucleation of the O2 phase relative to the energy penalty of increasing the oxidation state of the Ni ions in the P2 phase further than $Ni^{3+}$. Though a small amount of sodium ions may be further removed from the P2 phase with charge compensation through oxygen redox chemistry,[36] the Ni ions become inactive in the P2 structure at x<1/3.

In summary, we combined operando x-ray absorption spectroscopy with operando resonant elastic x-ray scattering to study a P2-$Na_xNi_{1/3}Mn_{2/3}O_2$ positive electrode material in a fully functional electrochemical coin cell. We directly measured the oxidation state change of Ni from $Ni^{2+}$ to $Ni^{3+}$ in the P2-phase and linked the $Ni^{3+}$ to $Ni^{4+}$ linked to the P2-O2 transition. We emphasize the importance of the operando characterization used here, because the highly charged O2 phase with $Ni^{4+}$ is likely unstable for a careful ex-situ characterization. Our work enables direct access to the correlation between local redox reactions and long-range order. It therefore opens new avenues to investigate underlying mechanisms within a wide variety of systems with structural phase changes and strain gradients that couple to redox activity, including sodium-ion and multivalent intercalation,[2,12,37,38] catalytic reactions [4,39], and corrosion.[40]

## Methods

**Synthesis/coin cell preparation.** To synthesize NNMO, transition metal nitrates $Ni(NO_3)_2 \cdot 6H_2O$ and $Mn(NO_3)_2 \cdot 4H_2O$ were titrated into a solution of stoichiometric NaOH at 10 ml/hr. Co-precipitated $M(OH)_2$ was filtered with a centrifuge and washed 3 times with DI water. The dried precursors were mixed and ground with stoichiometric $Na_2CO_3$ and calcined at 500 °C for 5 hr and at 900 °C for 14 hr in air. The cathodes were assembled by mixing NNMO particles with 10 wt% acetylene black and 5 wt% PTFE. 1 M $NaPF_6$ in 67 vol% diethylene carbonate and 33 vol% ethylene carbonate was the electrolyte, glass fiber was used as a separator and the counter electrode was Na metal, rolled thin to allow for x-ray penetration. The coin cells were assembled in an argon filled glovebox, with small holes drilled through both sides of the case to allow x-ray transmission. The holes were covered with Kapton tape and secured with small amounts of epoxy prior to coin cell assembly.

**Experimental details.** Cells were mounted on a 3D-printed sample holder and the experiment was conducted at the A2 beamline at the Cornell High Energy Synchrotron Source (Cornell University, Ithaca, NY). The cells were charged to 4.5 V at a constant-current rate of C/10 (approximately 0.1 mA, depending on the active mass of the cathode which was 5.5 and 6.0 mg for the two samples). X-ray data was collected from the directly transmitted beam by a diode as well as using a 2-D Pilatus detector to measure the (002) reflection of the cathode material. Non resonant data was collected using a photon energy of 15 keV and a $2\theta$ value of 9.08°, and at this energy the 2D detector could detect both the P2 (002) peak and the O2 peak when it appeared later in the charge. For this data, theta rocking curves of 200 points over 2 degrees with 2 seconds per point were collected.

For the resonant data, the photon energy was scanned repeatedly from 8.32 to 8.4 keV in 31 points, with a theta rocking curve of 21 images with 2 seconds per point captured at each energy. A diode was used to collect the directly transmitted intensity to measure absorption and a 2D Pilatus detector was used to record a powder diffraction ring corresponding to the (002) reflection of the NNMO particles ($2\theta = 15.18°$). 4 different locations in the cell, each separated by ~100 um, were cycled (each was sampled every 4 energy scans) to minimize x-ray damage to the probed material.

**Analysis.** Once collected, the data was averaged over each theta rocking curve. For the resonant data, as the $2\theta$ value changed with energy, the position of the reflection was manually adjusted so that the position was the same for all data as for the data collected at 8.32 keV (15.18°) (see Fig. S2a). The Debye-Scherrer rings appearing as arcs on the 2D detector were then flattened manually by estimating the arc as a parabola and shifting the ring to the $2\theta$ value of the ring at the center of the detector. The data was then averaged horizontally across the detector (see Fig. S2b). Background fluorescence was subtracted by averaging a small region of the detector at low $2\theta$ far below the peak and at high $2\theta$ far above the peak and doing a linear subtraction between those 2 regions for all detector images (see Fig. S3).

To calculate the absorption spectra, the ratio of the diode signal to the signal picked up by an ion chamber upstream from the sample was calculated and subtracted from 1 for all 31 energies in each energy scan throughout charging. To calculate REXS spectra, the total intensity of the (002) peak was plotted for each energy for each energy scan (see Fig. S4). Spectra were then rescaled such that the first point (at 8.32 keV) and the extrema at the edge (maximum for XAS and minimum for REXS) went from 0 to 1.

To calculate the energy of the onset of the absorption (resonant scattering) edges, a threshold of halfway between the signal at 8.32 keV and the maximum (minimum) was set. The spectra were linearly interpolated, and the lowest energy to have a signal greater than (less than) that threshold was chosen to be the energy of the edge onset (see Figure S5). Figure S6b shows calculated values of REXS spectra for the (002) peak of NNMO for decreasing concentrations of Na without changing the oxidation state of the Ni atoms present. The intensity of the spectra at energies above the intensity minimum decrease with decreasing Na like the measured spectra shown in Figure S6c, but at lower energies, no change occurs. The shift of the edge seen in the real spectra must therefore be due to the changing Ni oxidation state, not the Na concentration.

# Figures

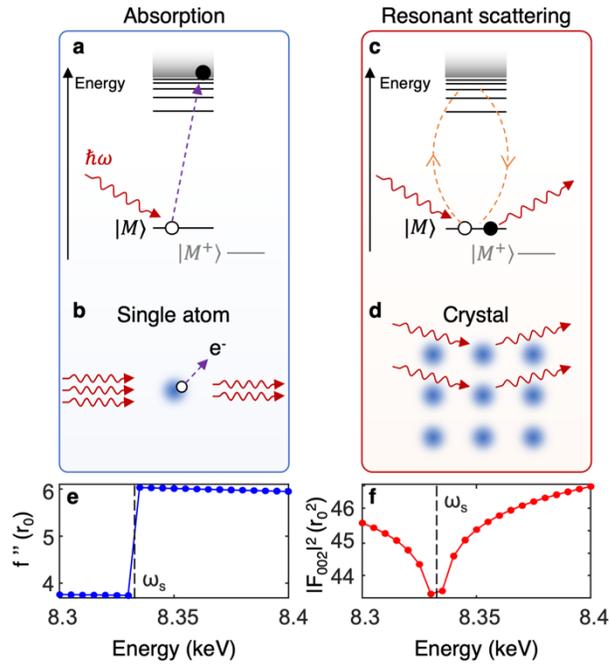

**Figure 1: Operando resonant elastic x-ray scattering.** (a) Schematic energy diagram of photoelectric absorption. (b) Schematic real space diagram of photoelectric absorption. A photon is absorbed by the atom and a core electron which is promoted to a higher unoccupied state or continuum. (c) Schematic energy diagram of resonant elastic x-ray scattering. (d) Schematic real space diagram of resonant scattering in a crystal. Photons interact with the crystal and are reemitted in phase, creating discrete intensity peaks per Bragg's Law. (e) Calculated imaginary dispersion correction f'', proportional to the absorption cross section, as a function of the photon energy around the Ni-K edge for a free-standing Ni-atom. f'' is energy independent until a discontinuous jump at the resonant frequency $\omega_s$. (f) Calculated energy spectrum of the intensity of the (002) Bragg reflection of the pristine P2-$Na_xNi_{1/3}Mn_{2/3}O_2$ calculated using the crystal structure and the dispersion corrections for a free-standing Ni-atom. The intensity reaches a minimum value at the same resonant frequency $\omega_s$ as the discontinuous jump in f''.

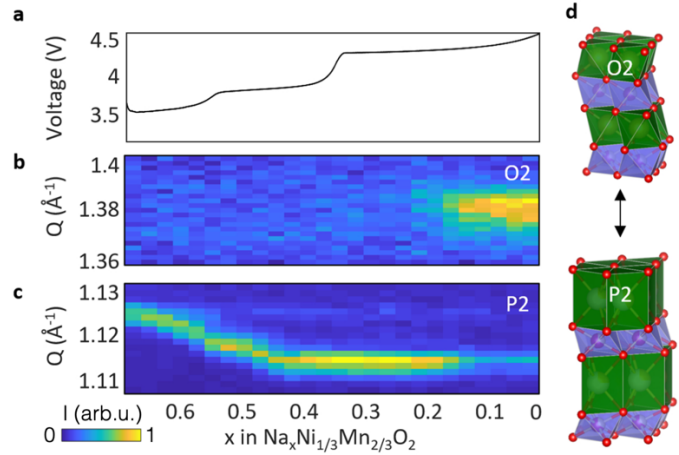

**Figure 2: Non-resonant operando x-ray diffraction on P2-Na$_x$Ni$_{1/3}$Mn$_{2/3}$O$_2$.** (a) Voltage profile during the duration of charging a cell to 4.5 V at a rate of C/10. (b,c) Diffraction data corresponding to the O2 and P2 crystal structures aligned to the same composition axis (equivalent to time in a constant charge-rate experiment). Diffraction data was recorded non-resonantly with a photon energy of 15 keV (λ = 0.827 Å). (d) Schematics of the structures of the O2 and P2 phases. A transformation between the two structures occurs through a shift of adjacent MO$_6$ layers relative to each other and a collapse of the layer spacing.

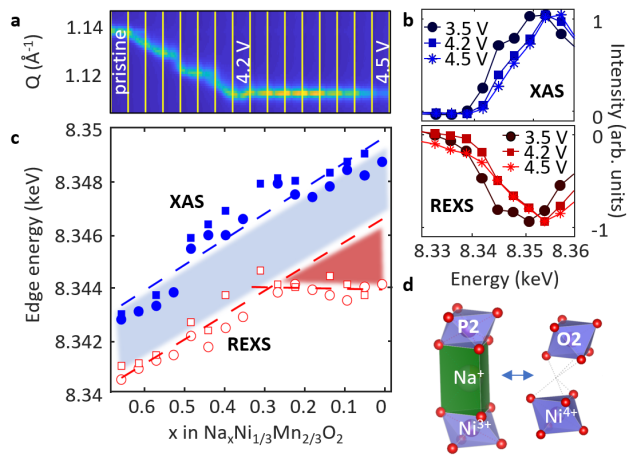

**Figure 3: Operando resonant elastic x-ray scattering.** (a) Intensity of the (002) peak of the P2 structure during charging recorded resonantly at the Ni K-edge. Vertical yellow lines indicate the beginnings/ends of energy scans. Conditions of pristine, 4.2 V, and 4.5 V are marked, and the false colors are identical to Fig. 2c. (b) Sample absorption and diffraction spectra at the marked times in Fig. (3a). (c) Energies of the absorption and diffraction edges (as calculated as the half max of the leading edge) during charging, aligned with Fig. (3a). Blue symbols represent absorption and red represent diffraction. Circles and squares represent separate trials. (d) Oxidation states of Ni shown in P2 and O2 phases at the transition marked by the plateau in Fig. (3c).

**Acknowledgements**

We thank Xinran Feng for assistance in the experiment, and Hector Abruña for lending a potentiostat for electrochemical measurements. The work at Cornell was supported by the National Science Foundation under Award # CAREER DMR-1944907 (operando x-ray measurements) and by the Center for Alkaline-based Energy Solutions, an Energy Frontier Research Center funded by DOE, Office of Science, BES under Award # DE-SC0019445 (method development and analysis of the x-ray data). The work at UC San Diego was supported by the National Science Foundation (NSF) under Award Number DMR1608968. Research conducted at CHESS was supported by the National Science Foundation under awards DMR-1332208 and DMR-1829070.

# Supplementary Information

## Structure-selective operando x-ray spectroscopy

*Daniel Weinstock[1], Hayley S. Hirsh[2], Oleg Yu. Gorobtsov[1], Minghao Zhang[2], Jason Huang[1], Ryan Bouck[1], Jacob P. C. Ruff[3], Y. Shirley Meng[2], and Andrej Singer[1]*

[1]*Department of Materials Science and Engineering, Cornell University, Ithaca, NY.*

[2]*Department of NanoEngineering, University of California San Diego, La Jolla, CA.*

[3]*Cornell High Energy Synchrotron Source, Cornell University, Ithaca, NY.*

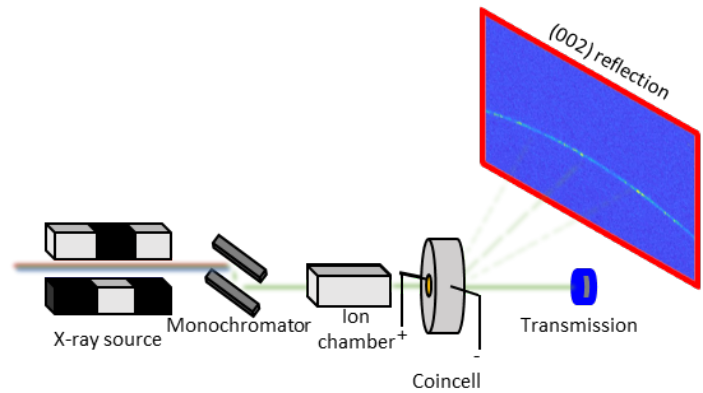

Figure S4: Schematic experimental setup of resonant x-ray measurements. The synchrotron-fed undulator and monochromator provide a highly tunable monochromatic photon beam that is swept over the Ni-K absorption edge from 8.32 to 8.4 keV. An ion chamber immediately upstream to the operando sample measures the intensity of the beam before interacting with the sample, allowing for accurate calculations of absorption and diffraction intensities. The coin cell is attached to a potentiostat that charges the cell at a rate of C/10 to 4.5 V with a constant current. Downstream of the sample are two detectors: a diode in line with the direct beam to measure transmission and a 2D detector positioned at $2\theta = 15°$ to measure the (002) Bragg reflection of the P2 phase.

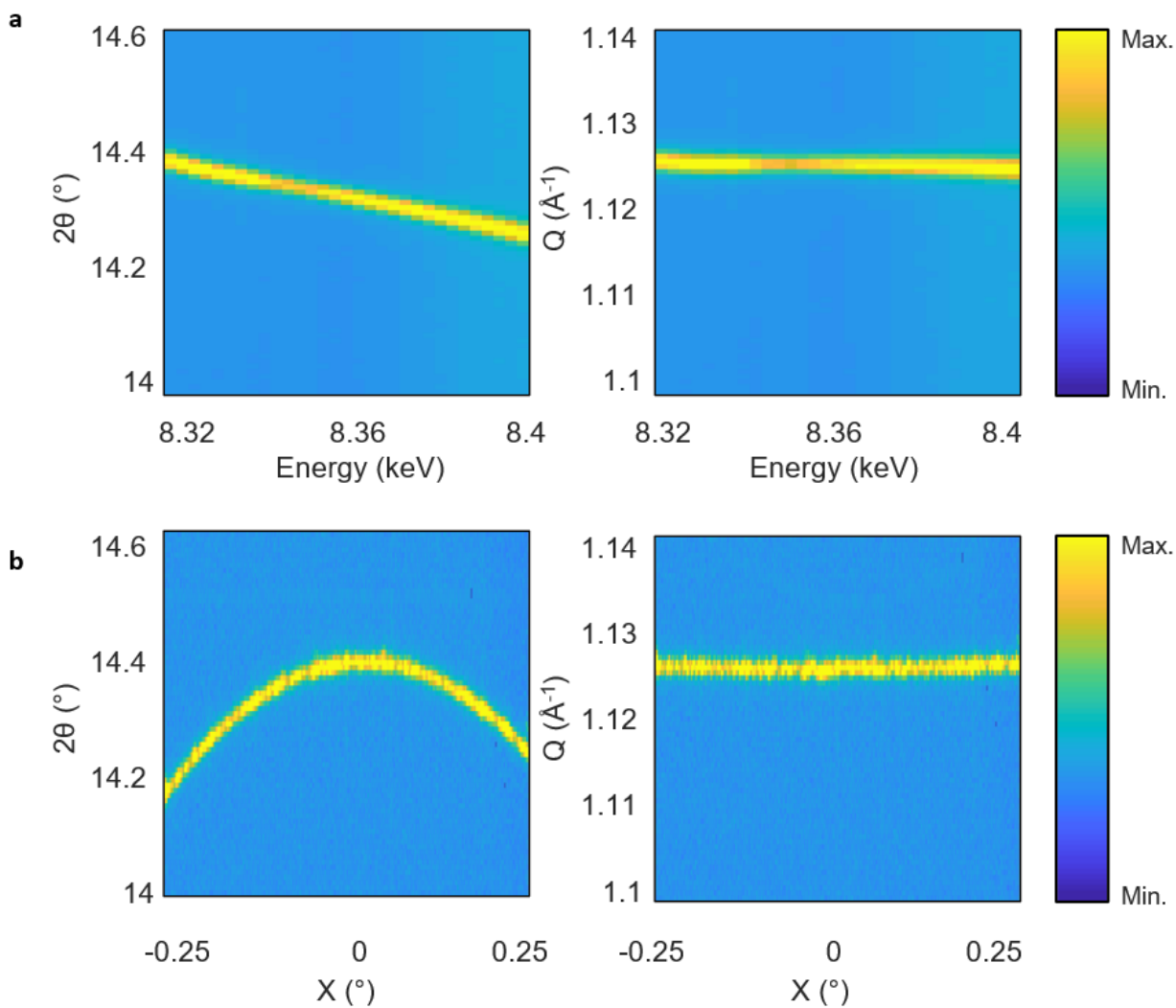

Figure S5: Demonstrations of calculations made during data analysis. **a.** Image of Q plotted against energy showing the shift in 2θ resulting from the changing energy (left) and corrected image (right). **b.** Detector image showing circular arc of the (002) peak (left) and flattened image (right). Images shown in linear scale.

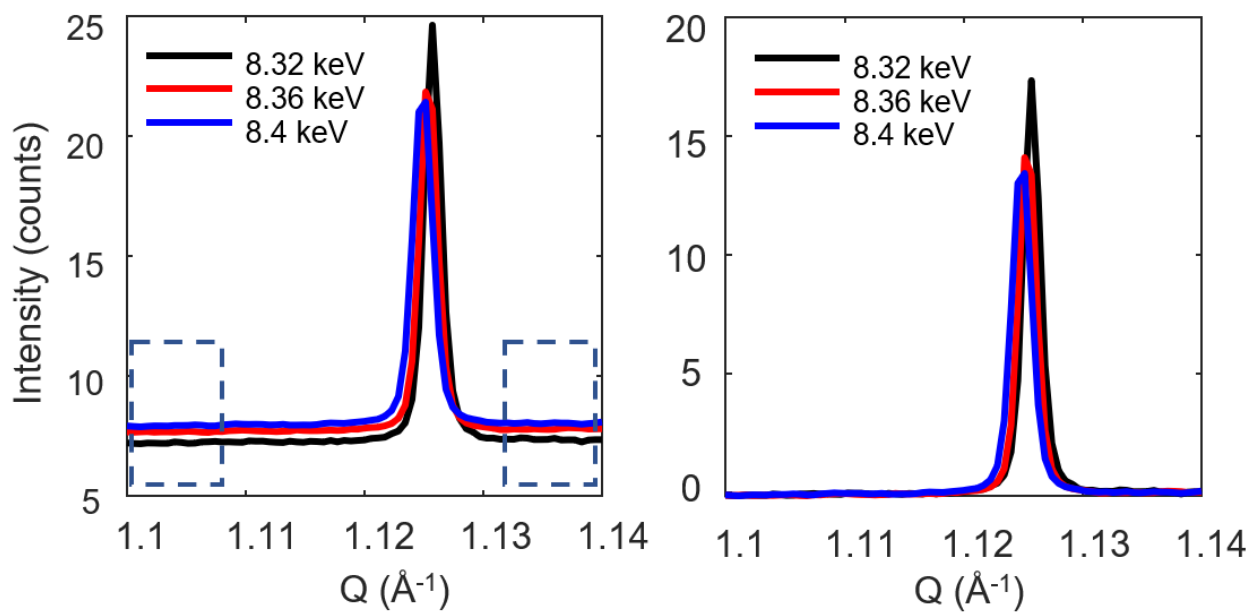

Figure S6: fluorescence removal shown for one energy scan. The intensity is the average number of photon counts per pixel along the detector perpendicular to Q. For each energy within each scan, the adjusted detector images (see Figure S2) were projected onto the Q axis (left). An average fluorescence signal was calculated by linear interpolation between the boxed areas away from any diffraction signal, and this signal was removed (right).

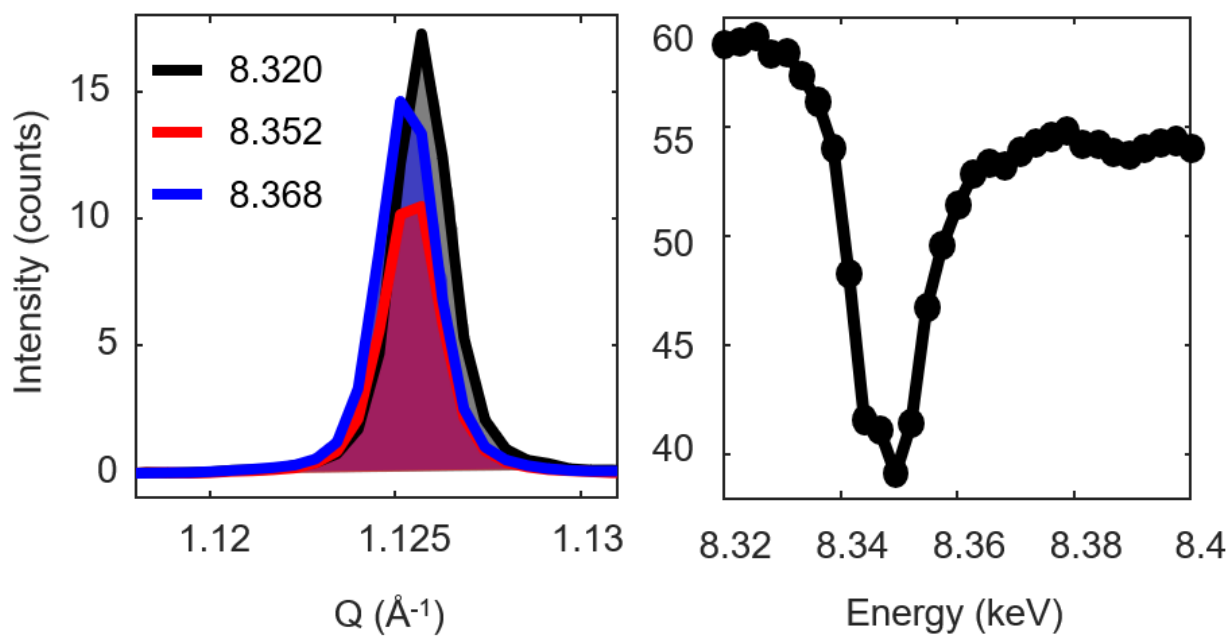

Figure S7: Calculation of REXS spectra from fluorescence-removed diffraction data (see Figure S3). For each energy at each scan, the diffraction data was integrated (right).

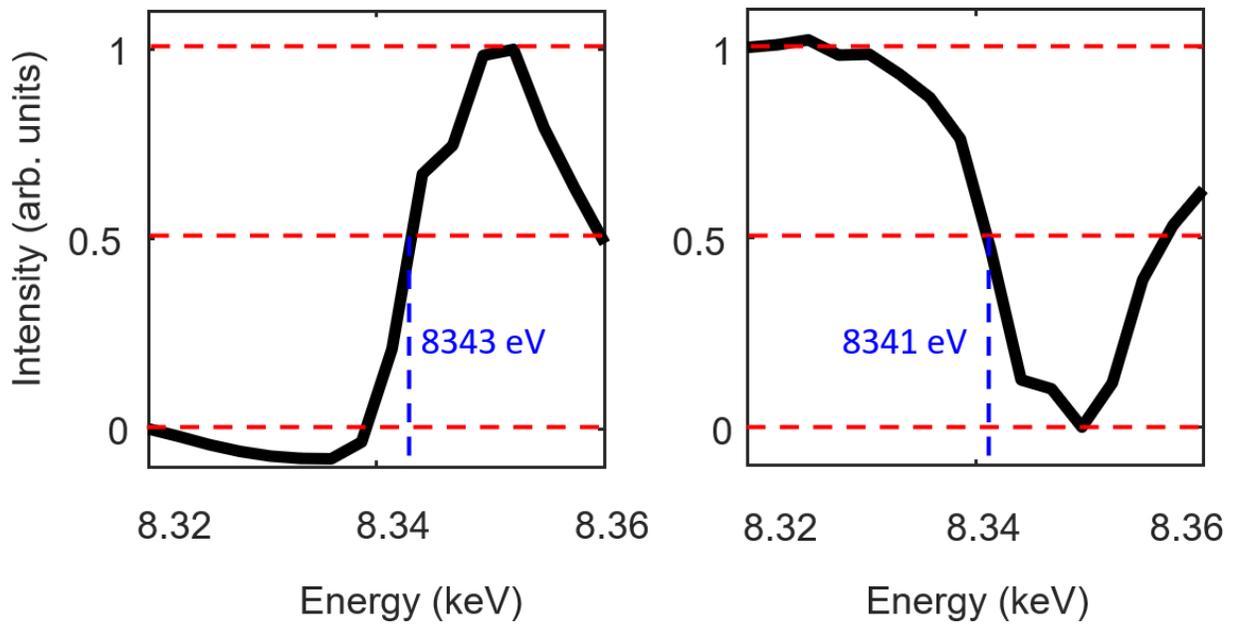

Figure S8: Calculation of edge energy from normalized XAS (left) and REXS (right) spectra.

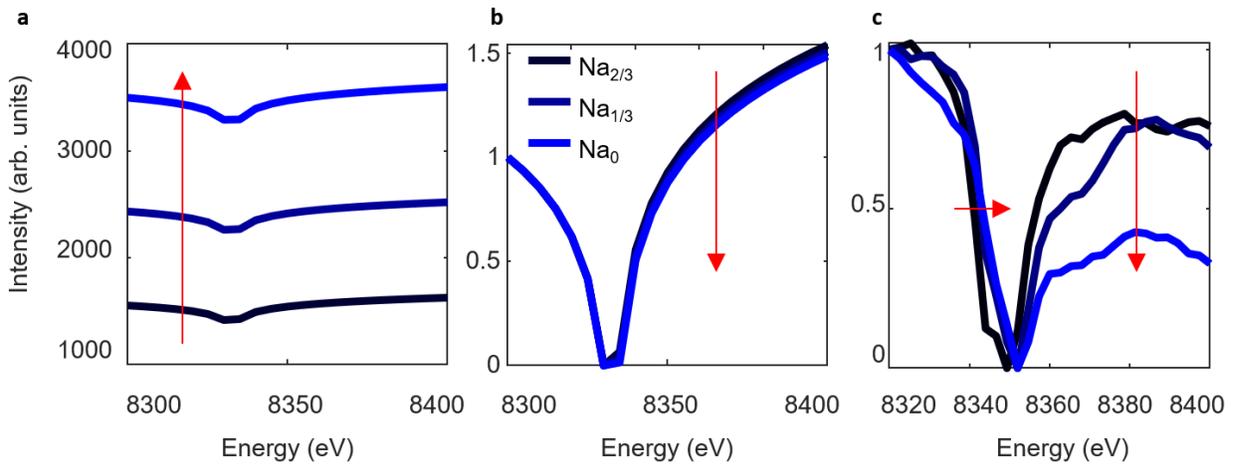

Figure S9: a. Simulated REXS spectra for the (002) peak of NNMO, with the Na concentration manually changed without adjusting the Ni oxidation state. Note the overall intensity increase with decreasing Na content, mirroring the increase in intensity seen in Figure 2c and 3a before the O2 phase begins to form. b. Data from Fig. S6 a, normalized. Note that right of the minimum the intensity drops with decreasing Na concentration, but left of the minimum, no change occurs with changing Na concentration – the shift in edge location only occurs with changing Ni oxidation state. c. Measured REXS spectra of the (002) peak of NNMO with the same Na concentrations of the calculated spectra in Fig. S6a. The real data shows a rightward shift of the dip left of the minimum, and a decrease in intensity after the minimum with decreasing Na concentration resulting from charging.

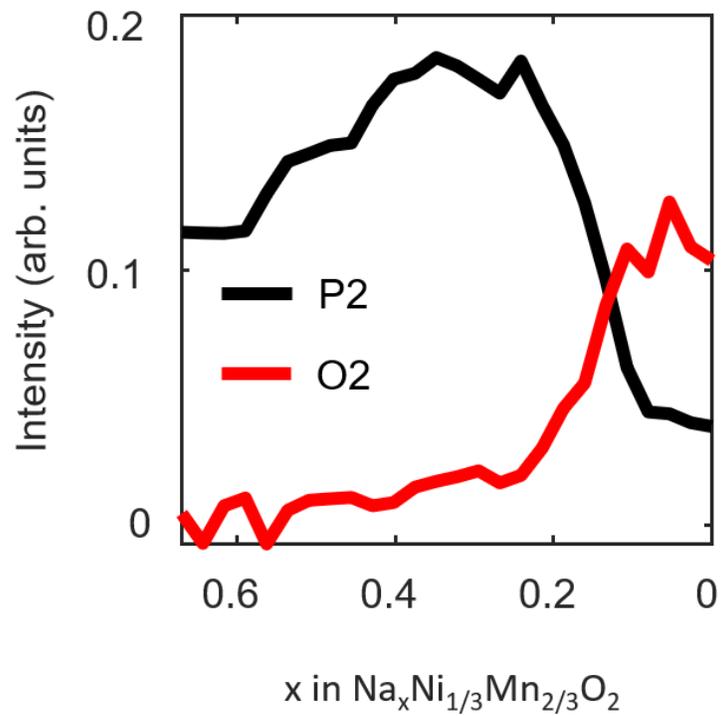

Figure S10: Integrated intensity of the P2 and O2 (002) peaks from the non-resonant x-ray measurements. The intensity of the P2 peak starts decreasing at approximately x=1/3, which is the onset of the P2-O2 phase transition.